\documentclass[aps,prc,groupedaddress,showpacs]{revtex4}
\usepackage[dvips]{graphicx}
\usepackage{epsfig}
\usepackage {amsmath}
\usepackage{amssymb}
\usepackage{float}
\include{epsf}

\def\iso#1#2{\mbox{${}^{#2}{\rm #1}$}}
\def\he#1{\iso{He}{#1}}
\def\li#1{\iso{Li}{#1}}
\def\be#1{\iso{Be}{#1}}
\def\b#1{\iso{B}{#1}}

\newcommand\beq{\begin{equation}}
\newcommand\eeq{\end{equation}}
\newcommand\beqar{\begin{eqnarray}}
\newcommand\eeqar{\end{eqnarray}}
\def\fun#1#2{\lower3.6pt\vbox{\baselineskip0pt\lineskip.9pt
  \ialign{$\mathsurround=0pt#1\hfil##\hfil$\crcr#2\crcr\sim\crcr}}}

\begin{document}

\title{Evaluation of Modern $^3$He($\alpha,\gamma)^7$Be Data}
\author{R. H. Cyburt}
\email[]{cyburt@nscl.msu.edu}
\affiliation{Joint Institute for Nuclear Astrophysics (JINA) and National Superconducting Cyclotron Laboratory (NSCL), Michigan State University, East Lansing, MI 48824}
\author{B. Davids}
\affiliation{TRIUMF, 4004 Wesbrook Mall, Vancouver BC V6T 2A3, Canada}

\begin{abstract}
In both the Sun and the early universe, the \he3($\alpha$,$\gamma$)\be7
reaction plays a key role.  The rate of this reaction is the least
certain nuclear input needed to calculate both the primordial \li7
abundance in big bang nucleosynthesis (BBN) and the solar neutrino
flux.  Taking advantage of several recent highly precise experiments,
we analyse modern \he3($\alpha$,$\gamma$)\be7 data using a robust and
minimally model dependent approach capable of handling discrepant data sets dominated by systematic rather than statistical errors. We find
$S_{34}(0)=0.580\pm0.043(0.054)$~keV~b at the 68.3(95.4)\% confidence level.
\end{abstract}


\maketitle

\section{Introduction}
\label{sect:introduction}

As the nearest star, the Sun is the best studied.  Models
predict the neutrino fluxes produced by the radioactive decay of
\be7 and \b8 in the solar core \cite{bahcall2}.  Over the last decade, various
neutrino observatories have measured the flux and flavour composition of the
\b8 neutrinos from the Sun ~\cite{Super-K,SNO}, allowing
constraints to be placed on neutrino mixing angles and mass-squared
differences.  New experiments sensitive to the
\be7 flux~\cite{borexino} are ongoing.  Both the \be7 and
\b8 neutrino flux predictions are nearly directly proportional to the astrophysical $S$ factor for the
$\he3(\alpha,\gamma)\be7$ reaction, $S_{34},$ at a relative
energy of $\sim 20$~keV; in fact, $\phi_\nu(\be7)\propto
S_{34}(0)^{0.86}$ and $\phi_\nu(\b8)\propto
S_{34}(0)^{0.81}$~\cite{bahcall1,adelberger}. This sensitivity allowed $S_{34}$ to be constrained using the measured solar neutrino fluxes and laboratory measurements of $S_{17}$, the astrophysical $S$ factor for the $\be7(p,\gamma)^8$B reaction \cite{cfo4}.  The
\he3($\alpha$,$\gamma$)\be7 reaction is important not only in
determining the solar neutrino fluxes, but also in other astrophysical
environs.

Primordial nucleosynthesis has been a robust prediction of hot big
bang cosmology for over 40 years~\cite{wagoner67,wssok,osw,fo06,fs06}.
The theory explains the large universal abundance of \he4 as well as
the origin of trace quantities of the light isotopes D, \he3, and
\li7.  It is a theory with three free parameters, the cosmic baryon
density, the neutron mean lifetime, and the number of active light
neutrino species. Via fits to the standard electroweak theory, measurements at the Large Electron Positron Collider have determined the number of active light neutrino species to be $N_\nu=2.984\pm0.008$~\cite{pdg08}, justifying the choice $N_\nu = 3$.  Many
experiments have been performed to determine the mean lifetime of the
neutron; the current Particle Data Group recommendation is $\tau_n =
885.7\pm0.8$ sec~\cite{pdg08}.  The cosmic baryon density has been
determined by analysing anisotropies in the cosmic microwave
background radiation; the latest WMAP satellite results are
$\Omega_bh^2=0.02273\pm0.00062$~\cite{wmap5}, where $\Omega_b$ is the universal mean density of baryons expressed in units of the critical density required to close the universe and the Hubble
constant is $100h$~km~sec$^{-1}$~Mpc$^{-1}$.  Using the Friedmann-Robertson-Walker cosmological model, the standard model of particle physics, and nuclear reaction rates one can therefore predict the light element abundances very precisely and compare directly with primordial abundance estimates based on observations of the oldest systems.

When observations of the Li abundances in the oldest stars in the Milky Way were compared with the predictions of BBN, a 2-3$\sigma$ discrepancy was found \cite{cyburt04}. Several possibilities exist for resolving this discrepancy, including particle physics beyond the standard model, improvements to the stellar models used to interpret astronomical observations, better approximations of the matter distribution in relativistic cosmology \cite{wiltshire07,leith08}, modifications of gravitational theory (e.g., \cite{bekenstein04}), and improved nuclear reaction rates. In this paper, we examine the last possibility. Lithium is made as beryllium in the early universe via the $\he3(\alpha,\gamma)\be7$ reaction.  Of course the rates of reactions that destroy \be7 must be known in addition to those that create it. Recent studies of the $\be7(d,p)2\alpha$ reaction cross section suggest that this reaction is not the source of the discrepancy \cite{coc04,angulo05}. The primordial \li7 abundance prediction is nearly directly proportional to the $\he3(\alpha,\gamma)\be7$ cross section at a relative energy of $\sim 300$~keV; the primordial abundance ratio \li7/H $\propto S_{34}^{0.96}$~\cite{cyburt04}.  On account of the
obvious importance of this reaction for both solar neutrinos and BBN, it has been subjected to
extensive theoretical as well as experimental study.

\begin{figure}
\includegraphics[width=\linewidth]{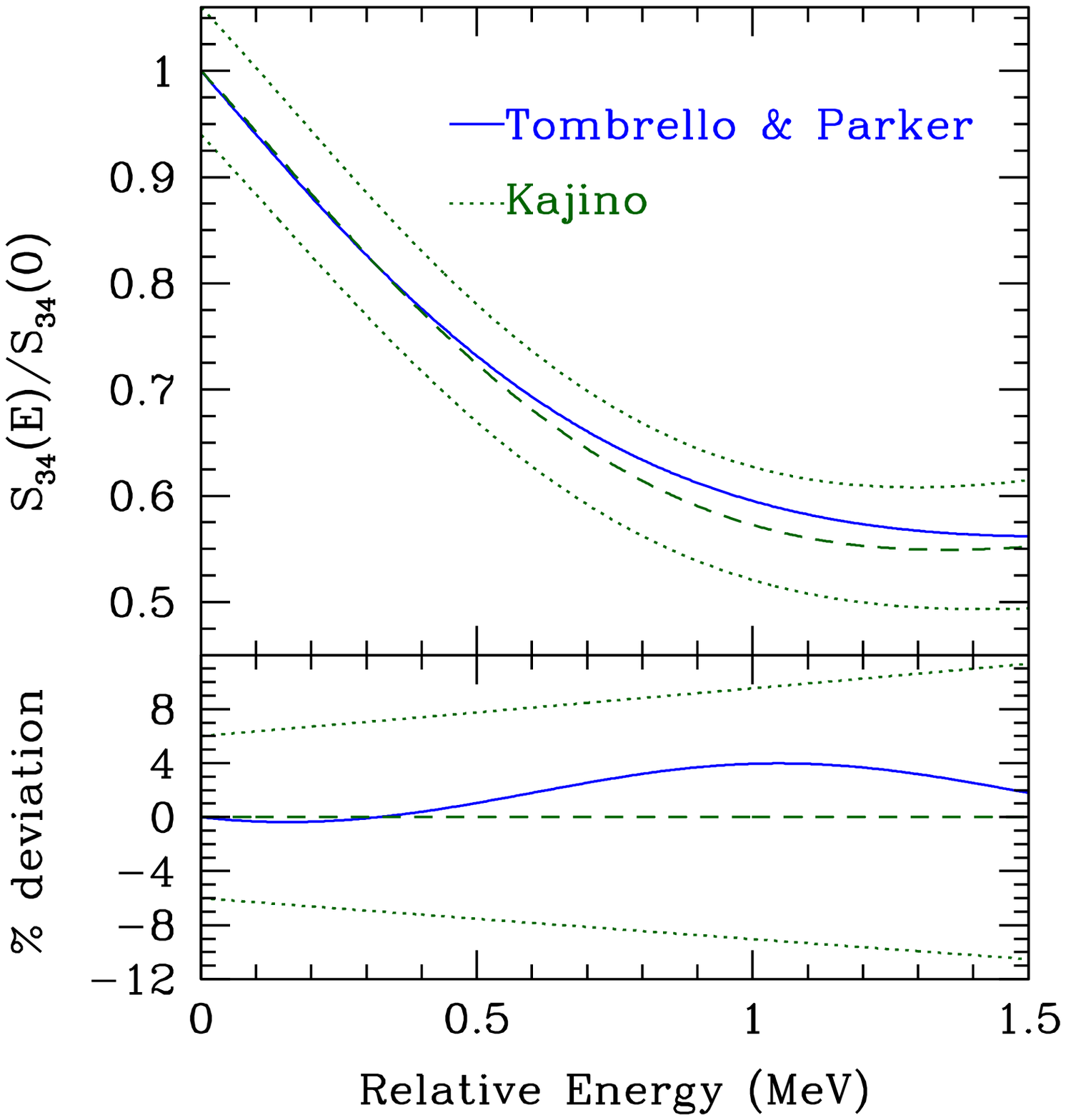}
\caption{Theoretical $S_{34}$ calculations by Tombrello and Parker~\cite{tombrello63} 
(solid) and by Kajino~\cite{kajino87} (dashed) are plotted relative to their
values at $E=0$ in the upper graph.  The deviations
relative to the Kajino calculation are plotted in the lower graph. The dotted lines delineate the uncertainty due to the $\pm$ 6\% theoretical uncertainty in the zero-energy $S$ factor and logarithmic derivative estimated in Ref. \cite{kajino86}.}
\label{fig:thry}
\end{figure}

Since the first potential model calculations
\cite{christy61,tombrello63}, many theoretical studies of the
\he3($\alpha,\gamma$)\be7 reaction have been performed
\cite{kajino84,buck85,kajino86,langanke86,kajino87,mohr93,igmanov97,csoto00,baye00}.  Examining the potential models, a hard sphere potential yields 
$S_{34}(0)=0.47$ keV b and $S^\prime(0)/S(0)=-0.60$
MeV$^{-1}$~\cite{tombrello63}, while a more physical potential yields
$S_{34}(0)=0.46$ keV b and $S^\prime(0)/S(0)=-0.79$
MeV$^{-1}$~\cite{baye00}.  Though the zero-energy $S$ factor values
are in good agreement, the shapes  as measured by the logarithmic derivatives are quite different. Similarly, in the single channel resonating group method (RGM) calculations of Ref.\ \cite{kajino86}, $S(0)$ varies widely with the nucleon-nucleon potential. The most commonly used fitting function for experimental data is the microscopic cluster model calculation of Kajino \cite{kajino84,kajino86,kajino87}. Ref.\ \cite{kajino86} reports $S_{34}(0)=0.50\pm0.03$ keV~b and $S^\prime(0)/S(0)=-0.548\pm0.033$ MeV$^{-1}$.  The estimated theoretical uncertainty of $\pm$6\%~\cite{kajino86} on both of these quantities must be considered when fitting this predicted shape to experimental data. The commonly cited potential model and microscopic cluster model calculations are shown in Figure \ref{fig:thry}. Another RGM calculation that includes the \li6+p configuration in addition to the \he3+$\alpha$ configuration \cite{csoto00} finds that the energy dependence of the astrophysical $S$ factor is not uniquely determined and yields a range of $S^\prime(0)/S(0)$ from $-0.70$ to $-0.50$ MeV$^{-1}$. Analyses that renormalize theoretically calculated $S$ factors must include an additional systematic error in order to account for uncertainties in the true shape of the $S$ factor when extrapolating to zero energy. It would be desirable to use the data themselves to determine the shape of the $S$ factor.

In this paper, we evaluate the modern prompt capture $\gamma$ ray and induced
\be7 activity measurements of the  $\he3(\alpha,\gamma)\be7$ cross section in a nearly model-independent way.  We use well known physics to constrain the energy dependence of the low energy cross section and disuss the treatment of systematic errors and their propagation into the final uncertainties, providing reliable values for $S_{34}$ at energies relevant to solar neutrino production and BBN with realistic uncertainties derived from Markov Chain Monte Carlo calculations.

\section{Modern Data}
\label{sect:data}

In the past 50 years, there have been many experimental efforts to measure the $\he3(\alpha,\gamma)\be7$ cross section, including those of Holmgren and Johnston~\cite{holmgren}, Parker and Kavanagh~\cite{pk63}, Nagatani {\em et al.}~\cite{nagatani}, Kr\"{a}winkel {\em et
al.}~\cite{krawinkel}, Robertson {\em et al.}~\cite{robertson}, Volk {\em et al.}~\cite{volk}, Alexander {\em et al.}~\cite{alexander}, Osborne {\em et al}~\cite{osborne}, and Hilgemeier {\em et al.}~\cite{hilgemeier}.   With the exception of Ref.\ \cite{holmgren}, these measurements were all considered in the Adelberger {\em et al.} review \cite{adelberger}, which recommended $S_{34}(0) = 0.53\pm0.05$ keV~b. This review noted a 2.5$\sigma$ discrepancy between the weighted averages of the measurements based on prompt $\gamma$ ray detection and those based on induced \be7 activity. At the time of this evaluation, we have the benefit of several subsequent, independent \be7 activity measurements as well as two prompt $\gamma$ ray detection studies. These new measurements employed improved detectors, better background suppression, and underground accelerators. Moreover, they were published almost two decades after the last of the measurements considered in Ref.\ \cite{adelberger}. We opt therefore to restrict this analysis to modern experimental data, namely  the recent measurements of Bemmerer {\em et al.} \cite{bemmerer06}, Brown {\em et al.} \cite{brown07}, Confortola {\em et al.} \cite{confortola07}, Gy\"{u}rky {\em et al.} \cite{gyurky07}, and Nara Singh {\em et al.} \cite{singh04}. As the data of Ref.\ \cite{bemmerer06} are included in the data of Ref.\  \cite{gyurky07}, we will refer henceforth only to the latter. The data considered here are shown in Table \ref{tab:data}.

\begin{table}
\caption{The modern data used in this evaluation.  Shown are the $S$ factors for capture into the ground ($S_0$) and first excited state ($S_1$) individually, or the sum of the two contributions ($S_{tot}$). Relative systematic errors for prompt measurements ($\delta_p$), activity measurements ($\delta_a$), and those that are common to the two methods ($\delta_c$) are shown separately.} 
\label{tab:data}
\begin{tabular}{lll}
\hline\hline
& Brown {\em et al.}~\cite{brown07} & \\
$\delta_p=0.0387$\footnote{This combines the 3.5\% systematic uncertainty with the 1\% varying systematic and 1.3\% energy uncertainties.} & $\delta_a=0.030$ & $\delta_c=0.027$ \\
& Prompt & \\
$E=0.3274$ & $S_0^{(p)}=0.349\pm0.012$ & $S_1^{(p)}=0.143\pm0.007$ \\
$E=0.4260$ & $S_0^{(p)}=0.311\pm0.004$ & $S_1^{(p)}=0.126\pm0.002$ \\
$E=0.5180$ & $S_0^{(p)}=0.302\pm0.005$ & $S_1^{(p)}=0.119\pm0.002$ \\
$E=0.5815$ & $S_0^{(p)}=0.280\pm0.007$ & $S_1^{(p)}=0.118\pm0.002$ \\
$E=0.7024$ & $S_0^{(p)}=0.268\pm0.005$ & $S_1^{(p)}=0.114\pm0.002$ \\
$E=0.7968$ & $S_0^{(p)}=0.260\pm0.002$ & $S_1^{(p)}=0.111\pm0.001$ \\
$E=1.2337$ & $S_0^{(p)}=0.227\pm0.002$ & $S_1^{(p)}=0.100\pm0.001$ \\
$E=1.2347$ & $S_0^{(p)}=0.236\pm0.002$ & $S_1^{(p)}=0.104\pm0.001$ \\
& Activity & \\
$E=0.3274$ & $S_{tot}^{(a)}=0.495\pm0.015$ & \\
$E=0.4260$ & $S_{tot}^{(a)}=0.458\pm0.010$ & \\
$E=0.5180$ & $S_{tot}^{(a)}=0.440\pm0.010$ & \\
$E=0.5815$ & $S_{tot}^{(a)}=0.400\pm0.011$ & \\
$E=0.7024$ & $S_{tot}^{(a)}=0.375\pm0.010$ & \\
$E=0.7968$ & $S_{tot}^{(a)}=0.363\pm0.007$ & \\
$E=1.2337$ & $S_{tot}^{(a)}=0.330\pm0.006$ & \\
$E=1.2347$ & $S_{tot}^{(a)}=0.324\pm0.006$ & \\ \hline
& Confortola {\em et al.}~\cite{confortola07} & \\
$\delta_p=0.038$ & $\delta_a=0.032$ & $\delta_c=0.023$ \\
& Prompt & \\
$E=0.0933$ & $S_0^{(p)}=0.3819\pm0.0170$ & $S_1^{(p)}=0.1451\pm0.0058$ \\
$E=0.1061$ & $S_0^{(p)}=0.3661\pm0.0132$ & $S_1^{(p)}=0.1519\pm0.0046$ \\
$E=0.1701$ & $S_0^{(p)}=0.3599\pm0.0076$ & $S_1^{(p)}=0.1501\pm0.0026$ \\
&  Activity & \\
$E=0.0929$ & $S_{tot}^{(a)}=0.534\pm0.016$ & \\
$E=0.1057$ & $S_{tot}^{(a)}=0.493\pm0.015$ & \\
$E=0.1695$ & $S_{tot}^{(a)}=0.507\pm0.010$ & \\ \hline
& Gy\"{u}rky {\em et al.}~\cite{gyurky07} & \\
& $\delta_a=0.031$ & \\
& Activity & \\
$E=0.1056$ & $S_{tot}^{(a)}=0.516(1\pm0.052)$ & \\ 
$E=0.1265$ & $S_{tot}^{(a)}=0.514(1\pm0.020)$ & \\ 
$E=0.1477$ & $S_{tot}^{(a)}=0.499(1\pm0.017)$ & \\ 
$E=0.1689$ & $S_{tot}^{(a)}=0.482(1\pm0.020)$ & \\ \hline
& Nara Singh {\em et al.}~\cite{singh04} & \\
& $\delta_a=0.0363$\footnote{The effective 
systematic error is derived from the energy dependent systematic errors 
with $\delta_{eff}^{-2}=\sum_i \delta_i^{-2}/N$.}& \\
& Activity & \\
$E=0.4200$ & $S_{tot}^{(a)}=0.420\pm0.014$ & \\
$E=0.5060$ & $S_{tot}^{(a)}=0.379\pm0.015$ & \\
$E=0.6145$ & $S_{tot}^{(a)}=0.362\pm0.010$ & \\
$E=0.9500$ & $S_{tot}^{(a)}=0.316\pm0.006$ & \\
\hline\hline
\end{tabular}
\end{table}

\section{Evaluation}
\label{sect:eval}

Several papers have discussed the energy dependence of nonresonant
low energy radiative capture cross sections \cite{christy61,jennings98,baye00}.  When reactions are dominated by external capture, the astrophysical $S$ factor exhibits a subthreshold pole
of the form
\beq
S(E) \propto \frac{1}{E+Q},
\eeq
where $E$ is the relative energy and $Q$ is the energy released by the reaction. This pole affects the convergence of the usual Maclaurin series expansions and
the shape of  cross sections at low energy, resulting in an upturn as $E\rightarrow0$.  If one wants better convergence, this subthreshold pole must be taken into account, expanding the quantity $(E+Q)S(E)$.

As shown in Ref.\ \cite{cyburt04a}, a completely model independent approach such as a Maclaurin series expansion does not work well when there are large discrepancies among data sets.  In such cases one requires a physical constraint on the shape of the $S$ factor. Since there have been many studies of the $\he3(\alpha,\gamma)\be7$ reaction, we can constrain the form of the fitting function beyond just a Maclaurin series using known physics. For instance, this radiative capture is dominated by $E1$ transitions at low energies \cite{tombrello63}.  Therefore, given the spin of nuclei involved, only $\ell=0$ and $\ell=2$ incoming partial waves contribute significantly to the external capture.

Following Ref.\ \cite{mukhamedzhanov02}, we expand the remaining terms accounting for Coulomb and pole effects and find
\beq
S(E) = \frac{Q}{E+Q} \left[ s_0(1+aE+\cdots)^2 + s_2(1+4\pi^2E/E_G)(1+16\pi^2E/E_G)(1+cE+\cdots)^2\right].
\eeq
Here, $s_0$ and $s_2$ are the amplitudes of the $s$ and $d$ wave $E1$
capture components respectively.  The terms with the
coefficients $a$ and $c$ are higher order contributions.  Keeping the high energy
behavior of these terms equivalent ($\propto E^2$ to lowest order)
demands $c\equiv0$ if $a$ is finite. $E_G$ is the Gamow energy given numerically by $E_G = 0.97913Z_1^2Z_2^2A$ MeV where $Z_1e$
and $Z_2e$ are the charges of the reactants and $A$ their
reduced mass in atomic mass units.  For $S_{34}$, $E_G=26.9437$ MeV.

The $\he3(\alpha,\gamma)\be7$ reaction can proceed via
capture into either of the spin-orbit partners, the \be7 ground state
($Q=1.5861$ MeV) and the first excited state at 429 keV ($Q=1.1570$ MeV).
When determining best fits and zero-energy extrapolations, one must
bear this in mind and use experiments sensitive to not only the total
capture cross section, but also the partial capture cross sections
into the ground and first excited states. We are therefore left with 3 parameters for the $S$ factor fits for each capture
(i.e. ground or excited state).

In general, modern experiments are dominated by systematic errors.  As
such, standard statistical treatments used to combine data from
different, possibly discrepant experiments are not valid.  The traditional method of scaling the error in the mean by $\sqrt{\chi^2/\nu}$, where $\nu$ is the number of degrees of freedom, has been shown to poorly estimate the error when systematics dominate the error budget \cite{d'agostini94,cfo1,cyburt04}.  Such treatments lead to an underestimation of the true uncertainties. We require a more reliable prescription for this analysis. 

If one assumes that the dominant systematic error is the overall normalization
error, several techniques can be used to estimate the true
uncertainty. It was shown in Ref.\ \cite{cyburt04} that a simple
robust measure of systematic uncertainty is determined by the weighted
dispersion of the fit.  In another approach, d'Agostini
\cite{d'agostini94} suggests floating the normalizations of individual
experiments weighted by their uncertainties as a way of compensating
for the shortcomings of the standard statistical approach.  This
method works well, and agrees with the error estimate of Ref.\
\cite{cyburt04}. In the modern era, the size of systematic errors is often greater than or equal to the size of statistical errors. Hence systematic errors must be treated properly in order to obtain a reliable central value and uncertainties .

We use a $\chi^2$ minimization procedure to determine the parameters of our calculated $S_{34}$ that best fit the data.  We break up the $\chi^2$ into two components, $\chi^2_{data}$ and $\chi^2_{norm}$, such that $\chi^2 = \chi^2_{data} +
\chi^2_{norm}$. Thus for a single data set $n$, 

\beqar
\chi^2_{data}(n) &=& \sum_i \left( \frac{S(E_i)-\alpha_nS_i}{\alpha_n\sigma_i} \right)^2 \textrm{and} \\
\chi^2_{norm}(n) &=& \left( \frac{\alpha_n-1}{\delta_n}\right)^2.
\eeqar
Here, $\alpha_n$ and $\delta_n$ are the floating renormalization factor
and the normalization uncertainty of the dataset $n$, $S(E_i)$ is the theoretically calculated S factor at $E_i$, $S_i$ is the measured $S$ factor at $E_i$, and $\sigma_i$ is the standard deviation of the measured $S$ factor at $E_i$.

In the case where two experiments have correlated normalizations, $\chi^2_{norm}$ is modified and becomes
\beq
\chi^2_{norm}(n1,n2) = \frac{\left[ \delta_{n2}^2(\alpha_{n1}-1)^2 - 2\delta_c^2(\alpha_{n1}-1)(\alpha_{n2}-1) + \delta_{n1}^2(\alpha_{n2}-1)^2 \right]}{\left[ \delta_{n1}^2\delta_{n2}^2 - \delta_c^4 \right]}
\eeq
where $\delta_{n1}$ and $\delta_{n2}$ are the total systematic errors
of the data sets $n1$ and $n2$ and $\delta_c$ is the systematic error
common to both data sets.  This equation is the result of inverting the
covariance matrix.  

We use a Markov Chain Monte Carlo (MCMC) approach
to evaluate the data, following the two-step procedure laid out by
Ando {\em et al.} \cite{ando06}.  First, we find the best fit by varying the
parameters randomly, with ever decreasing step sizes until the results converge
to a predetermined number of significant figures.  Then, with step sizes
determined by $\Delta\chi^2=1$ variations, we start a random walk away from the best fit point using the Metropolis algorithm~\cite{metropolis,MCmeth}.  Once the sample variance of each of the
parameters converges within the specified resolution, we stop the MCMC.  We tested convergence with chains of length up to $10^8$, finding convergence after about $10^6$ steps.  We use the $10^8$ step chain for our final results.

\section{Results}
\label{sect:results}

To test the validity of the adopted three parameter $S_{34}$ model, we found the best
fits for both two and four parameter models as well,  fixing $a=0$ for
the two parameter fit and letting both $a$ and $c$ vary for the
four parameter fit.  We find best fits with $\chi^2_{tot} = 85, 63, 61$
or $\chi^2_{tot}/\nu = 2.74, 2.17, 2.26$ for the two, three, and four parameter fits, respectively.  One can see readily that given the modern data considered here, we can not adequately determine more than 3 parameters. Also, the two parameter fit is clearly unrepresentative of the data, whilst the quality of the four parameter fit is no better than that of the
three parameter fit. Hence including a fourth parameter is unwarranted.  The uniqueness of the solution
was tested by starting the minimization procedure from different parameter values, finding the same minimum.

After the best fit is found by minimizing $\chi^2$ and the MCMC
has converged, we find the limits of central confidence intervals by determining
parameter values that obey the relation $P(\nu/2,\Delta\chi^2/2) =
CL$~\cite{press92}. Here $P(a,x)=\gamma(a,x)/\Gamma(a)=1-\Gamma(a,x)/\Gamma(a)$ is the regularized incomplete gamma function~\cite{temme96}, $\nu=N_{par}+N_{norm}$ is the total number of varied parameters and normalizations, $\Delta\chi^2=\chi^2-\chi^2_{min}$,
and $CL$ is the desired confidence level.  
\begin{table}
\label{tab:results}
\begin{tabular}{l|ll|l}
\hline\hline
Parameter & Mode & Mean $\pm$ Std. Dev. & Norm. Error \\
\hline
$\alpha_p$(Brown)      & 0.95 & $0.95\pm0.02$ & 0.0387  \\
$\alpha_a$(Brown)      & 0.95 & $0.95\pm0.02$ & 0.030  \\
$\alpha_p$(Confortola) & 0.99 & $0.99\pm0.02$ & 0.038  \\
$\alpha_a$(Confortola) & 1.01 & $1.01\pm0.02$ & 0.032  \\
$\alpha_a$(Gy\"{u}rky) & 1.02 & $1.02\pm0.02$ & 0.031  \\
$\alpha_a$(Nara Singh) & 1.04 & $1.04\pm0.02$ & 0.0363 \\
\hline
$s_0$(gs)            & 0.406   & $0.406\pm0.009$ &\\
$s_2$(gs)            & 0.007   & $0.007\pm0.001$ &\\
$a$(gs)              & -0.207  & -$0.203\pm0.038$ &\\
\hline
$s_0$(ex)            & 0.163   & $0.163\pm0.004$ &\\
$s_2$(ex)            & 0.004   & $0.004\pm0.001$ &\\
$a$(ex)              & -0.134  & -$0.127\pm0.055$ &\\
\hline
$\chi^2_{norm}$        & 4.83  & $6.48\pm2.02$ & $N_{norm}=6$\\
$\chi^2_{data}$        & 58.38 & $69.06\pm4.70$ &$N_{data}=41$\\
\hline
$\chi^2_{tot}$         & 63.21 & $75.54\pm4.95$ &$N_{par}=6$  \\
\hline\hline
\end{tabular}
\caption{Results of the Markov Chain Monte Carlo parameter estimation.} 
\end{table}

\begin{figure}
\includegraphics[width=\linewidth]{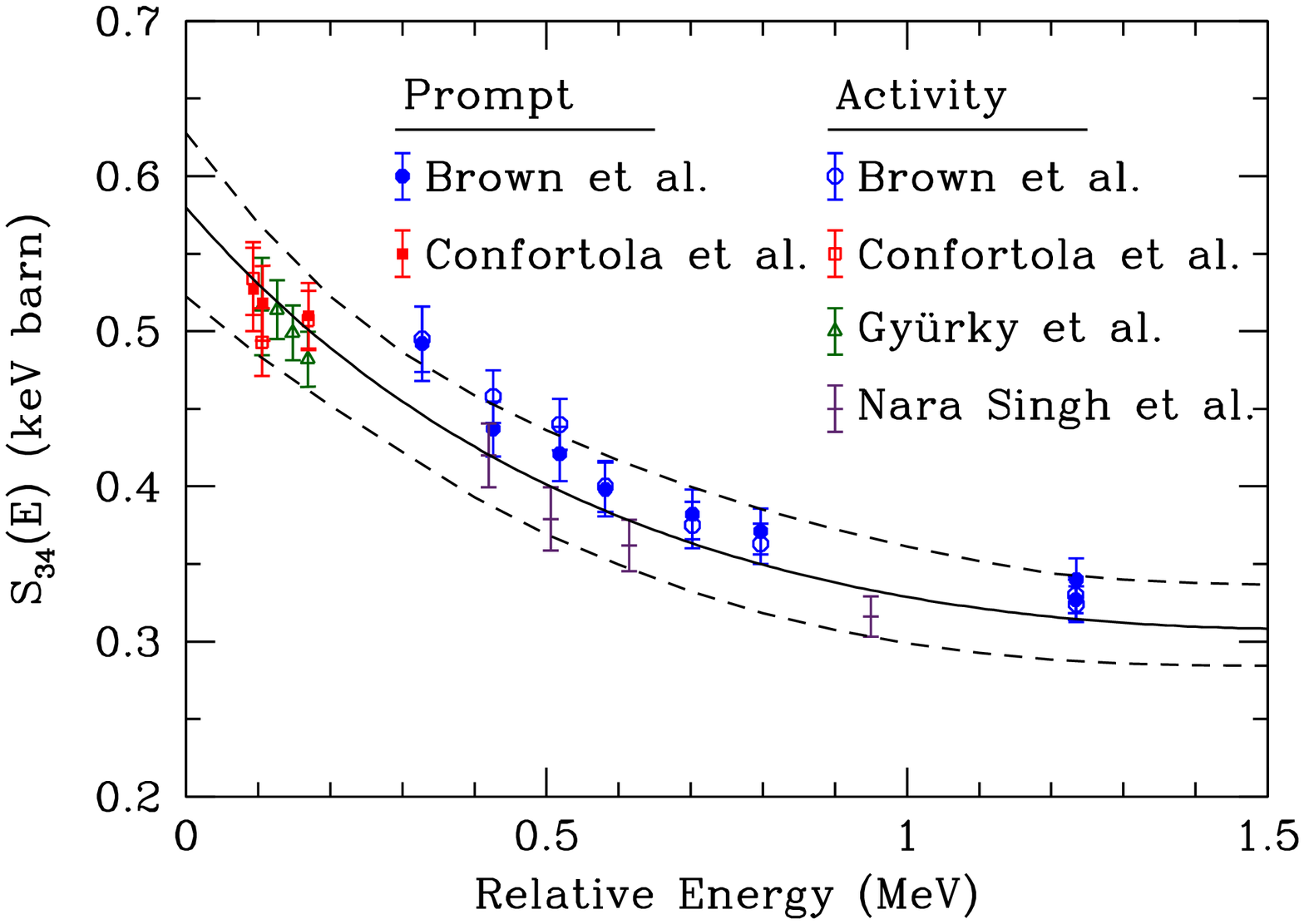}
\caption{(Color online) The best fit $S$ factor for the \he3($\alpha$,$\gamma$)\be7 reaction
(solid curve).  Also shown are the central 68.3\% confidence level limits (dashed
curves).  The experimental data and their total uncertainties are plotted (Activity-open points, Prompt-solid points).}
\label{fig:S34}
\end{figure}

The results of this analysis are presented in Table \setcounter{table}{1}
II and Figure \ref{fig:S34}. We find a mode and central 68.3\% CL interval of $S_{34}(0) = 0.580\pm0.043$ keV b ($\Delta S_{34}(0)/S_{34}(0)= 7.4$\%). We find a mean and standard deviation of $S_{34}(0) = 0.580\pm0.013$ keV b
($\Delta S_{34}(0)/S_{34}(0)=2.2$\%).  The small size of the interval given by the mean $\pm$ the standard deviation compared to the central 68.3\% CL results from marginalizing over all parameter
distributions.  The standard deviation here corresponds to defining errors
with $\Delta\chi^2=1$, known to underestimate uncertainties when there
is more than 1 degree of freedom. The central 95.4\% confidence interval is $S_{34}(0)=0.580\pm0.054$ keV~b. The size of this relative uncertainty is slightly smaller than that given in the evaluation of Adelberger {\em et al.} \cite{adelberger}, which was a 1$\sigma$ error. Hence the modern data considered here permit a considerably more precise recommendation for $S_{34}$(0), even when allowing the shape of the $S$ factor to be determined by the data rather than a theoretical model and when fitting the branching ratio as well as the total $S$ factor.

\begin{figure}
\includegraphics[width=\linewidth]{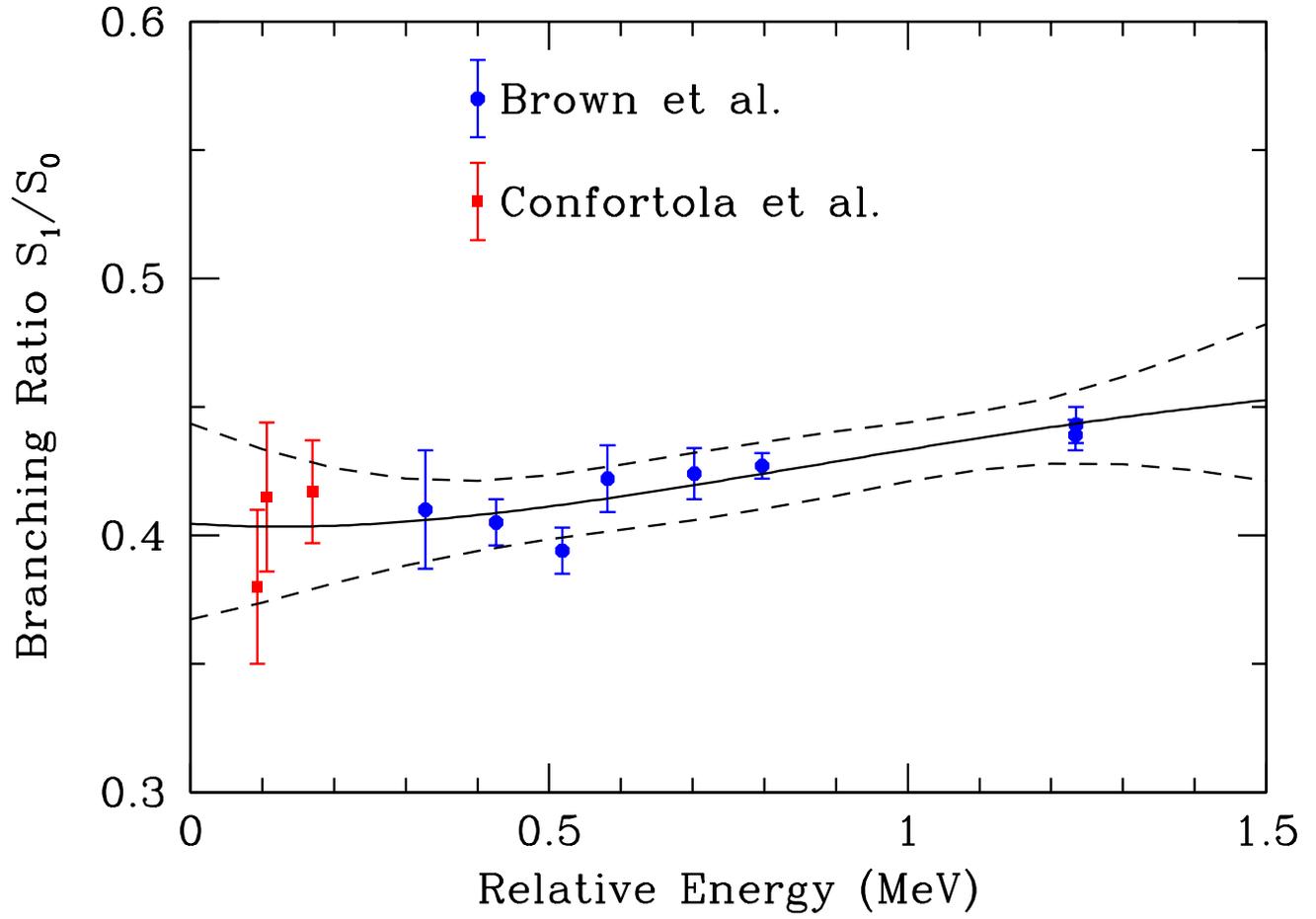}
\caption{(Color online) The best fit branching ratio $S_1/S_0$ for the
\he3($\alpha$,$\gamma$)\be7 reaction (solid curve).  Also shown are the
central 68.3\% confidence level limits (dashed curves).  The experimental
data and their total uncertainties are plotted.}
\label{fig:R34}
\end{figure}

\begin{figure}
\includegraphics[width=\linewidth]{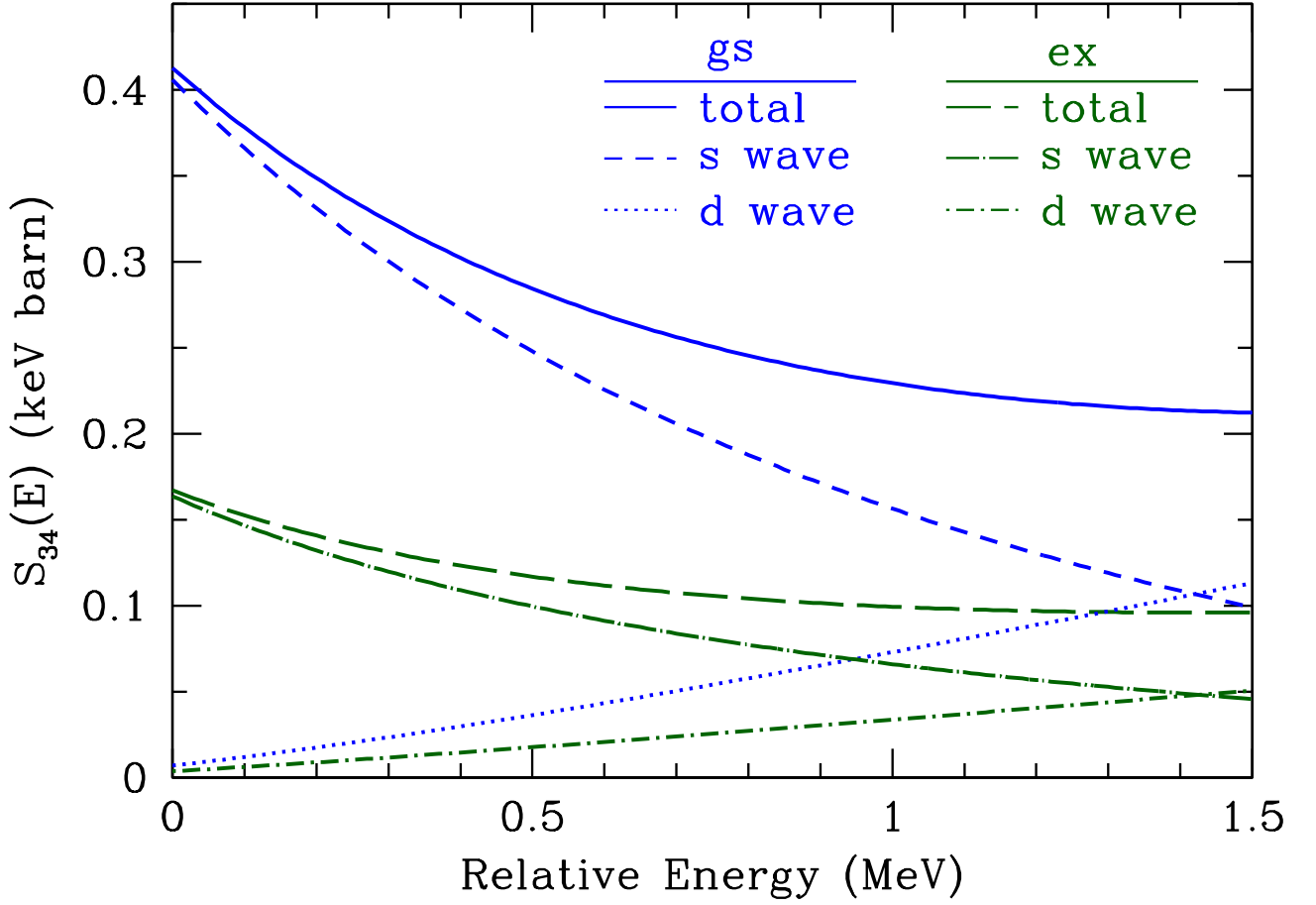}
\caption{(Color online) This figure shows the fitted contributions of various partial waves to the $S$ factor.  The s-wave (short dashed), d-wave (dotted) and total (solid) $S$ factors for capture into the ground state and the s-wave (long dash-dotted), d-wave (short dash-dotted) and total (long dashed) $S$ factors for capture into the first excited state are all shown.}
\label{fig:comp}
\end{figure}

Using the best fit parameters and the definitions in~\cite{cyburt04}, we find a discrepancy error of
5.3\% and a normalization error of 3.0\%, yielding a total 
systematic uncertainty of 6.1\%, in very good agreement with the present MCMC results,
considering that this error does not take into account discrepancies in
the branching ratio $S_1/S_0$, as the MCMC method does. The MCMC results for the branching ratio are shown in Figure \ref{fig:R34}. The s- and d-wave contributions to the ground and excited state transitions determined in the MCMC analysis are shown in Figure \ref{fig:comp}.

We have calculated the thermally averaged rate of the \he3($\alpha$,$\gamma$)\be7 reaction per particle pair $\lambda_{34}$ and fit it using the form
\beq
\lambda_{34} = \exp{\left[ a_0 + a_2/T_9^{1/3} + a_6\ln{T_9} \right]}
\frac{\left( 1 + n_1T_9^{2/3} + n_2T_9^{4/3} \right)}{\left( 1 +
d_1T_9^{2/3} + d_2T_9^{4/3} \right)} \textrm{cm$^{3}$ mol$^{-1}$ s$^{-1}$}.
\eeq
The fit parameters are shown in Table III. This functional form provides a better fit to the reaction rate than the fitting functions of Caughlan and Fowler \cite{cf88} and REACLIB \cite{sakharuk06,smith08} and
agrees with the numerically calculated rate within 0.5\%.

\setcounter{table}{2}
\begin{table}
\label{tab:rate}
\begin{tabular}{l|l|l|l}
\hline\hline
Parameter & Low & Adopted & High \\
\hline
$a_0$ & 15.531721 & 15.609867 & 15.679639 \\
$n_1$ & -0.100208 & -0.020478 &  0.037757 \\
$n_2$ &  0.235187 &  0.211995 &  0.196645 \\
$d_1$ &  0.114322 &  0.255059 &  0.353050 \\
$d_2$ &  0.373802 &  0.338573 &  0.316019 \\
\hline\hline
\end{tabular}
\caption{Shown are the fit parameters for the
recommended thermal rate ($\lambda_{34}$) and its central 68.3\% confidence level limits, with
$a_2=-12.82707707$, $a_6=-2/3$, and
$a_0=\ln{[1.03762\times10^{7}S_{34}(0) \textrm{keV}^{-1} \textrm{b}^{-1}]}$ fixed.}
\end{table}
  
\begin{figure}
\includegraphics[width=\linewidth]{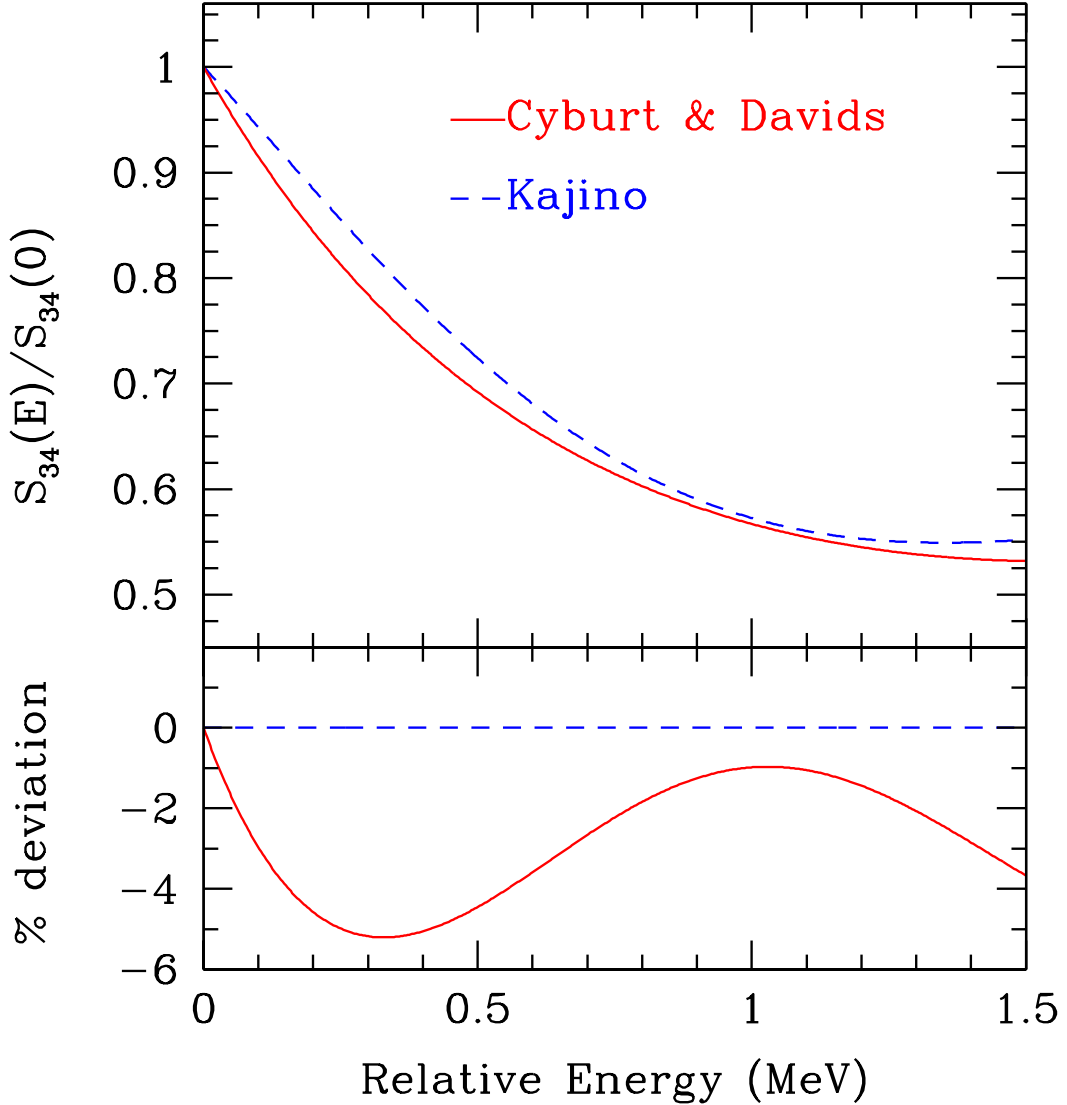}
\caption{\label{cdk}(Color online) Comparison of the shape of the RGM calculation of Kajino \cite{kajino87} with the present structure model-independent analysis.}
\end{figure}

Figure \ref{cdk} compares the energy dependence of $S_{34}$ determined here from a nearly model-independent analysis of modern data with that calculated by Kajino {\em et al.} \cite{kajino87}. It is apparent that in the energy range of principal interest below 500 keV, the shapes of the two curves differ substantially. Extrapolation using the RGM model of Kajino from the energy range important for BBN to that of interest in the Sun would differ by some 5\% from extrapolation based on the $S$ factor determined here. This is quite consistent with the estimates of the theoretical error given in Ref.\ \cite{kajino86}.

\clearpage

\section{Conclusions}
\label{sect:conclusions}

We have performed a minimally model dependent analysis of modern \he3($\alpha$,$\gamma$)\be7 data. This analysis properly takes into account the systematic errors of potentially discrepant data and yields reliable central values and uncertainties. At the 68.3\% confidence level, we find $S_{34}(0)=0.580\pm 0.043$ keV~b and $S^\prime(0)/S(0) = -0.92\pm0.18$ MeV$^{-1}$. We have used this value of $S_{34}(0)$ to compute the thermally averaged rate of the \he3($\alpha$,$\gamma$)\be7 reaction per particle pair, and have fit this rate with an analytical form accurate within 0.5\%.

This reaction rate is about 9\% higher than the recommendation of \cite{adelberger}, but is quite consistent with it. It is also perfectly consistent with the recommendations of References \cite{nacre,de04}. Moreover, access to precise modern data has allowed us to minimize the uncertainties associated with extrapolation and arrive at a more precise and better founded value that does not depend on nuclear structure models.

Our recommended reaction rate implies that the predicted \be7 and \b8 solar neutrino fluxes should both be increased by 8\% compared with predictions based on the $S_{34}(0)$ value from Ref.\ \cite{adelberger}. Similarly, the present analysis implies that the value of $S_{34}$ in the energy range relevant to big bang nucleosynthesis is 17\% larger than the value adopted in Ref.\ \cite{cyburt04}, resulting in a 17\% increase in the predicted primordial \li7 abundance. The uncertainties found here
reduce the total error in the \li7 abundance prediction by a factor of $\sim2$. The discrepancy between the predicted and observed primordial \li7 abundances increases from $\sim3\sigma$ to $\sim5\sigma$~\cite{cfo5}, implying a rather serious disagreement that must be resolved.  

In order to improve our knowledge of the rate of the \he3($\alpha$,$\gamma$)\be7 reaction, future experiments should collect data at at least six different energies. No modern data exist in the Gamow window for big bang nucleosynthesis, so this would be of primary interest. Information on capture into the ground and first excited states would also be useful, arguing in favour of prompt $\gamma$ ray measurements rather than those based on induced \be7 activity. Finally, precise measurements from 1 - 3 MeV would allow better constraints on the shape of the $S$ factor at the low energies of astrophysical interest.

\section{Acknowledgments}

RHC would like to thank T.A.D. Brown for useful discussions and providing $S$ factor data. RHC is supported through
NSF grant PHY 02 16783 (JINA) and BD acknowledges support from the Natural Sciences and Engineering Research Council of Canada. TRIUMF receives federal funding via a contribution agreement through the National Research Council of Canada. 

\bibliography{s34}

\end{document}